\documentclass[10pt,twocolumn,british,showpacs]{revtex4}
\usepackage[latin1]{inputenc}
\usepackage{color}
\usepackage{graphicx}

\makeatletter

\providecommand{\tabularnewline}{\\}

\usepackage{color}
\usepackage{graphicx}
\usepackage{amssymb}

\usepackage[hypertex]{hyperref}
\newcommand{\F}{Fig.~}

\usepackage{amsmath}
\usepackage{bm}
\renewcommand{\mathbf}{\bm}
\usepackage{ifthen}
\newboolean{includeNotes}
\setboolean{includeNotes}{false}

\usepackage{babel}

\makeatother
\begin{document}

\title{Binary fluids under steady shear in three dimensions}

\author{K. Stratford$^{1,2}$}

\author{J.-C. Desplat$^{3}$}


\author{P. Stansell$^{1}$}

\author{M. E. Cates$^{1}$}

\affiliation{$^{1}$SUPA, School of Physics and $^{2}$EPCC, The University of Edinburgh, JCMB The King's
Buildings, Mayfield Road, Edinburgh, EH9 3JZ, United Kingdom; $^{3}$Irish Centre
for High-End Computing, Dublin Institute for Advanced Studies, 5 Merrion Square,
Dublin 2, Ireland}

\begin{abstract}
We simulate by lattice Boltzmann the steady shearing of a binary fluid
mixture with full hydrodynamics in three dimensions. Contrary to some
theoretical scenarios, a dynamical steady state is attained with finite
correlation lengths in all three spatial directions. Using large
simulations we obtain at moderately high Reynolds numbers apparent scaling exponents
comparable to those found by us previously in 2D. However, in 3D there
may be a crossover to different behavior at low Reynolds number:
accessing this regime requires even larger computational resource
than used here. 
\end{abstract}

\pacs{{\footnotesize 64.75.+g, 47.11.Qr}}

\maketitle
Systems that are not in thermal equilibrium play a central role in modern statistical
physics \cite{SFM}. They include two important classes: those evolving
towards Boltzmann equilibrium (e.g., by phase separation following a temperature
quench), and those maintained in nonequilibrium by continuous driving (such
as a shear flow). Of fundamental interest, and surprising physical subtlety, are
systems combining both features --- such as a binary fluid undergoing phase separation
in the presence of shear. Here a central issue \cite{Onuki02,Cates99} is whether coarsening
continues indefinitely, as it does without shear, or whether a nonequilibrium steady state (NESS) is reached,
in which the characteristic length scales $L_{x,y,z}$ of the fluid domain structure
attain finite $\dot{\gamma}$-dependent values at late times. (We define the mean
velocity as $u_{x}=\dot{\gamma}y$ so that $x,y,z$ are velocity, velocity gradient
and vorticity directions respectively; $\dot{\gamma}$ is the shear rate.) 

Our recent simulations, building on earlier work of others \cite{WagnerYeomans99,Gonnella01}, have shown that in two dimensions (2D), a NESS is indeed achieved \cite{stansell}. In 3D, the situation is more subtle. Fourier components of the composition field whose wavevectors lie along the vorticity direction feel no direct effect of the mean advective velocity \cite{Onuki02,catesmilner}. Therefore it might be possible for coarsening to proceed indefinitely by pumping through tubes of fluid oriented along $z$ \cite{Cates99}. Another crucial difference is that in 2D fluid bicontinuity is possible only by fine tuning to a percolation threshold at 50:50 composition (assuming fluids of equal viscosity) so that the generic situation is one of droplets. (Indeed, for topological reasons, droplets are implicated even at threshold \cite{WagnerYeomans99}.) In contrast, in 3D both fluids remain continuously connected across the sample throughout a broad composition window either side of 50:50. 

In 3D experiments, saturating length scales are reportedly reached after a period of
anisotropic domain growth \cite{Hashimoto888994,Onuki02}. However, the extreme elongation
of domains along the flow direction means that, even in experiments, finite size
effects could play a role in such saturation \cite{Bray03}. Theories
in which the velocity does not fluctuate, but does advect the diffusive fluctuations
of the concentration field, predict instead indefinite coarsening, with length scales
$L_{y,z}$ scaling as $\dot{\gamma}$-independent powers of the time $t$ since quench,
and (typically) $L_{x}\sim\dot{\gamma}tL_{y}$ \cite{Bray03}. 
As emphasized in \cite{stansell}, in real fluids, however,
the velocity fluctuates nonlinearly in response to the advected concentration
field, and hydrodynamic scaling arguments, balancing interfacial and either viscous
or inertial effects, predict saturation instead e.g., $L/L_{0}\sim(\dot{\gamma} T_{0})^{-1}$
or $L/L_{0}\sim(\dot{\gamma} T_{0})^{-2/3}$ \cite{DoiM91,Onuki97,Cates99}.
Here, $L_{0} = \nu^2/(\rho\sigma)$ and $T_{0} = \nu^3/(\rho\sigma^2)$,
with $\rho$ density, $\nu = \eta/\rho$ kinematic viscosity
and $\sigma$ interfacial tension, are the characteristic length and time at which inertial effects start to influence coarsening \cite{Kendon01}. Given these uncertainties as to the fate of sheared binary fluids in 3D, computer simulations of such systems,
with full hydrodynamic velocity fluctuations, are of great interest. 

Such simulations also offer demanding challenges to the state of the art
in computational physics. 
The 2D lattice Boltzmann (LB) results of
\cite{stansell} were obtained from 16 production runs involving
lattices ranging from $512\times 256$ to $2048\times 1024$ (all systems
having aspect ratio 2:1). Many pre-production runs were required to steer
 simulation
parameters so as to avoid finite-size effects and other artefacts. This
effort was rewarded, however: the unique parametric flexibility of LB allowed us to probe over six decades of reduced shear rate $\dot\gamma T_{0}$ \cite{stansell}.
Below, we extend that work to three dimensions with $9$
production runs on $512 \times 256 \times 256$ lattices, and $3$
larger runs of $1024\times 512\times512$ (i.e., all with aspect ratio 2:1:1).
Even given the excellent parallel scaling of LB on multi-processor
machines, each one of these $12$ datasets
required more computational resource than the entirety of
Ref.\cite{stansell}. The production runs reported here were performed using
1024 processors of the IBM Blue Gene/L machine at the University of
Edinburgh.  

\begin{figure}
\includegraphics{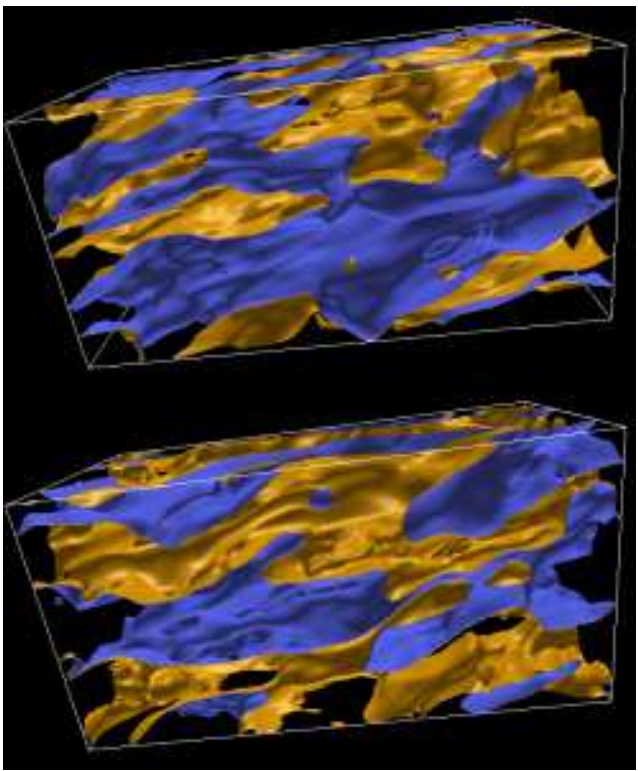}
\vspace{-0.4cm}
\caption{
Snapshots of the interface position at $\dot\gamma T_0 = 22.47$ (top) and $47.45$ (bottom) with parameter set R019 (Table 1). These are representative of the observed NESS. The mean flow is
rightward along the upper face of the simulation box and leftward at the lower face; the line of sight lies close to the vorticity (neutral) direction, $z$. Yellow and blue isosurfaces are constructed at
$\phi = \pm 0.2$  to create a dividing surface color-coded by the adjacent fluids (both shown transparent).\label{fig:PhiFields}}
\end{figure}

Although our simulations are not the first to address sheared binary fluids in 3D (see e.g. \cite{Cates99,harting}), earlier studies have offered only inconclusive evidence of NESS formation in systems free of finite size effects. Such effects can cause fully lamellar or hexagonal cylindrical domains, which wrap the periodic boundary conditions with simple topologies that prevent further hydrodynamic coarsening \cite{CatesPT,Cates99}; but this ``trivial'' route to NESS relies directly on the periodic boundary conditions and is thus not available in the bulk-system limit. Below we present evidence of NESS formation in systems retaining the complex topology expected in bulk samples, where a steady state dynamical balance can arise between the coarsening of bicontinuous domains under the action of interfacial tension, and their stretching by the flow (Fig.\ref{fig:PhiFields}). 

%

The required parameter steering would not have proven possible without
having the 2D runs to initially guide our selection --- a methodology
that can only succeed if the physics in 2D and 3D is not radically
different. Below we find that to be true for the upper few decades
of the range of $(\dot\gamma T_{0})^{-1}$ addressed in~\cite{stansell};
within this range, evidence is given below for saturation of correlation lengths with finite values in all three directions.
We then combine datasets using a quantitative scaling methodology
developed for the unsheared problem in \cite{Kendon01} and for shear
in \cite{stansell}; this allows scaling exponents to be estimated
using combined multi-decade fits. Caution is required here due to residual
finite size effects; these are unavoidable, particularly at high shear rates where we find NESS hardest to achieve numerically. Note that high shear rates correspond to {\em low} Reynolds numbers Re $ \simeq L_y^2\dot\gamma/\nu$
(due to the decrease of domain size with shear rate); these results could therefore signify new physics at low Re \cite{Cates99}. However, much larger systems sizes might
be needed to gain full access to this regime.

\begin{table*}
\begin{center}\begin{tabular}{ccccccccccccc}
\hline 
Name&
 $\nu$&
 $M$&
 $\sigma_{\textrm{theory}}$&
 $\sigma_{\textrm{meas}}$&
 $L_{0}$&
 $T_{0}$&
$\xi_0$&
$\dot{\gamma}$&
 $\Lambda_x$&
 $L_x$&
 $L_y$&
 $L_z$
\tabularnewline
\hline
R028&
 1.41&
 0.05&
 0.063&
 0.055&
 36.1&
 927&
 1.13&
5.0$\times 10^{-4}$&
1024&
--&
--&
--
\tabularnewline
R029&
 0.2&
 0.15&
 0.047&
 0.042&
 0.952&
 4.54&
1.13&
5.0$\times 10^{-4}$&
1024&
--&
--&
--
\tabularnewline
R020&
 0.025&
 2&
 0.0047&
 0.0042&
 0.149&
 0.886&
1.13&
5.0$\times 10^{-4}$&
1024&
--&
--&
--
\tabularnewline
R003&
0.015&
2.0&
0.0047&
0.0042&
0.054&
0.19&
1.13&
7.5$\times 10^{-4}$&
512&
511&
72.2&
172
\tabularnewline
&
&
&
&
&
&
&
&
5.0$\times 10^{-4}$&
512&
828&
116&
352
\tabularnewline
R004&
0.01&
2.0&
0.0047&
0.0042&
0.024&
0.0567&
1.13&
7.5$\times 10^{-4}$&
512&
356&
68.1&
131
\tabularnewline
&
&
&
&
&
&
&
&
5.0$\times 10^{-4}$&
512&
491&
106&
192
\tabularnewline
R030&
 0.00625 &
 1.25&
 0.0047&
 0.0042&
 0.00930&
 0.0138&
1.13&
5.0$\times 10^{-4}$&
512&
375&
91.6&
160
\tabularnewline
R007&
0.005&
2.0&
0.0047&
0.0042&
0.0059&
0.00709&
1.13&
5.0$\times 10^{-4}$&
512&
382&
97.4&
174
\tabularnewline
R008&
0.0035&
2.0&
0.0047&
0.0042&
0.0029&
0.00243&
1.35&
5.0$\times 10^{-4}$&
512&
370&
101&
177
\tabularnewline
R019&
 0.0014&
 4&
 0.0024&
 0.0021&
 0.000933&
 0.000622&
1.35&
5.0$\times 10^{-4}$&
512&
234&
71.3&
118
\tabularnewline
R032&
 0.0005&
 5&
 0.00094&
 0.00083&
 0.000301&
 0.000181&
1.35&
5.0$\times 10^{-4}$&
512&
135&
48.0&
71.2
 \tabularnewline
\hline
\end{tabular}
\end{center}

\caption{\label{tab:3d}
Parameter sets used in 3D simulations and observed NESS length scales.
Where a trivial NESS could be identified by inspection, no
length is recorded. The results of R020 were ambiguous:
periods of apparent NESS were contaminated by intervals of partial remixing (low
$\left< \phi^2 \right>$).
}
\end{table*}


The governing equations for our binary fluid system are the
Cahn-Hilliard equation for the composition $\phi$,
and the incompressible
Navier-Stokes equation for the velocity $u_{\alpha}$ in an isothermal
fluid of unit density $\rho$:
\begin{eqnarray}
\left(\partial_{t}u_{\alpha}+u_{\beta}\nabla_{\beta}u_{\alpha}\right)+\nabla_{\alpha}p-\nu\nabla^{2}u_{\alpha}-\phi\nabla_{\alpha}\mu & = & 0\label{two}\\
\partial_{t}\phi+\nabla_{\alpha}\left(\phi u_{\alpha}-M\nabla_{\alpha}\mu\right) & = & 0\label{three}
\end{eqnarray}
Here, $p$ is pressure (related in LB to density fluctuations, which
are small \cite{Kendon01});
$\nu$ is the kinematic viscosity;
$M$ is the  ($\phi$-independent) mobility and
$\mu=B\phi\left(\phi^{2}-1\right)-\kappa\nabla^{2}\phi$ is
the chemical potential. $B$ and $\kappa$ are positive constants;
the interfacial tension is $\sigma=(8\kappa B/9)^{1/2}$ and the
interfacial width is $\xi_{0}=(2\kappa/B)^{1/2}$ \cite{Kendon01}.

We solve these equations with an LB algorithm similar to 
that reported in \cite{Kendon01,Swift96}. 
To achieve the necessary shear rates, the domain is decomposed blockwise using multiple
Lees-Edwards sliding periodic boundary conditions \cite{stansell,Adhikari05},
chosen so that $\int_{0}^{\Lambda_{y}}\nabla_y u_{x}\, dy=\Lambda_{y}\dot{\gamma}$.
Although we neglect thermal fluctuations in our
fluid, as appropriate for dynamics near a zero-temperature fixed point
\cite{Bray94}, a fluctuating local velocity field still arises via
nonlinear interaction between the order parameter field and flow field. 
To help control errors, we adhered as far as possible to previously used parameter values and protocols \cite{Kendon01,stansell}. However, the sheared 3D case
showed significant stability problems compared with either the
3D unsheared case or 2D sheared case. To alleviate these, we replaced the D3Q15 lattice of \cite{Kendon01} with a D3Q19 model; this removes a ``computational mode'' responsible for some of the instabilities of D3Q15 \cite{d3q19}. We also use a multiple relaxation time
approach \cite{jstatphys} in place of a single
relaxation time \cite{Kendon01,stansell}, further improving stability.

Most of our 3D production runs were made using system size 
$512\times 256\times 256$,
run for $t\simeq 4\times10^{5}$ time steps.
Holding other parameters fixed, one finds that if $\dot{\gamma}$ is too
small, the domain size is large and finite size effects dominate, whereas if
$\dot{\gamma}$ is too large then the domains become small on the
lattice scale and tend to form a partially (or even fully) remixed state with 
strongly blurred interfaces. Such remixing could be a real physical effect at shear rates so high that the local interfacial structure departs strongly from equilibrium, but this happens at much lower shear rates in an LB fluid than in a real one (where  $\xi_0$ is much smaller). We therefore reject as artefacts all such partially remixed states, as identified by a significant reduction in order parameter
variance $\left<\phi^2\right>$. Worst affected were the runs at higher Reynolds number (low viscosity) where an adjustment
of the interfacial width from $\xi_0 =1.13$ to $\xi_0 = 1.35$ helped to
maintain acceptable behavior. All simulations reported here
were done for fully symmetric quenches with parameters
summarized in Table~1. As in the unsheared case
\cite{Kendon01}, judicious combinations of $\xi_{0}$, $\sigma$, $M$ and
$\nu$ allow systems spanning several decades in $L/L_{0}$ and
$\dot{\gamma}T_{0}$ to be accurately studied by varying $L_{0}$ and $T_{0}$ alongside $\dot{\gamma}$.

\begin{figure}
\includegraphics*{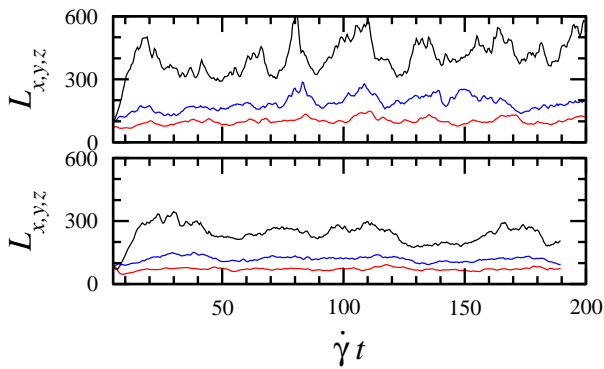}
\vspace{-0.4cm}

\caption{\label{fig:timeSeries} Two examples of $L_{x,y,z}$ in lattice units
as a function
of time in strain units $\dot{\gamma} t$ for R030 parameters (upper panel)
and R019 parameters (lower panel).
For all parameters where a steady state is observed, the length scale
as measured by the gradient statistic of \cite{WagnerYeomans99} is largest in the velocity
direction $L_x$, followed by the vorticity direction $L_z$, with that in
the velocity gradient direction $L_y$ the smallest.}
\end{figure}

\F\ref{fig:PhiFields} shows snapshots of the interfacial structure
based on the order parameter field for R019 with
$\dot{\gamma}=5\times10^{-4}$ after a steady state had been reached \cite{supplementary}.
\F\ref{fig:timeSeries} shows time series for $L_{x,y,z}$
from runs R030 and R019 as measured by a standard order parameter gradient
statistic \cite{WagnerYeomans99} that effectively measures
the mean distance between interfaces crossing the chosen direction. 

In \cite{Kendon01}, finite size effects (in the absence of shear) were considered {\em quantitatively} under control when the correlation length $L$ was less than 1/4 of the system size $\Lambda$. In \cite{stansell} this criterion was applied to time-averaged correlation lengths $L_{x,y}$ in the 2D sheared system. However, the actual system size dependence of $L_{x,y,z}$ in both 2D \cite{stansell} and 3D (this work) suggests that under shear this criterion is unnecessarily strict, at least if the purpose is to eliminate the {\em qualitatively} artefactual states that arise directly from finite size effects. As mentioned previously, these ``trivial'' NESS's form obvious laminar stripes extending the full size of the simulation box in both $x$ and $z$ directions. For such states, $L_{x,z}$ values that are formally much larger than the simulation dimensions $\Lambda_{x,z}$ are rapidly established. ($L_{x} \gg \Lambda_x$ means that that, for most coordinates $y,z$, one can cycle round the periodic boundary conditions in $x$ without encountering a single domain wall.)
To formally eliminate these, a criterion $L_{x,y,z} \le \Lambda_{x,y,z}$ is applied, which also excludes one apparently nontrivial NESS run (Table~1) from the scaling analysis made below. At the lowest Reynolds numbers investigated, only a trivial NESS was found on a $512\times 256\times 256$ lattice; larger
systems, $1024\times 512\times 512$, were then simulated for these parameters but gave the same structure. This difficulty in achieving bulk NESS at low Re perhaps suggests onset of a new regime; we return to this below.

\begin{figure}

\includegraphics*{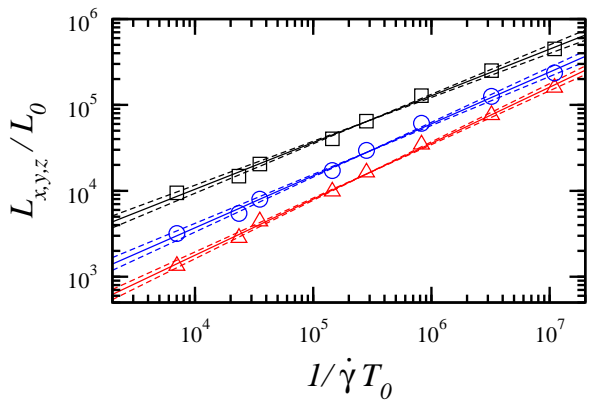}
\vspace{-0.4cm}

\caption{\label{fig:scalings}
Reduced length scales $L_{x,y,z}/L_{0}$ (black squares, red triangles, blue circles respectively) as a function of inverse
reduced shear rate for the 8 runs in which nontrivial NESS was observed.
The standard errors in the
individual points are no larger than the symbols; the dashed lines give the
95\% confidence limits of the fitted regression.}

\end{figure}

Only at the largest $(\dot\gamma T_0)^{-1}$ values investigated was the strict finite size criterion of \cite{Kendon01}, $L_x<\Lambda_x/4$, approached. (Note however that earlier studies accepted $L<\Lambda/2$ as sufficient, e.g. \cite{jury}.) Accordingly we expect that the {\em quantitative} scaling of all our correlation length data with shear rate may still be affected by finite size corrections. With this caveat, we proceed to perform a scaling analysis based on the protocol of \cite{stansell}.
To construct our scaling plot, mean values of $L_{x,y,z}$ were obtained
via a bootstrap procedure \cite{stansell} performed on each times series,
discarding data
for which $t<10^{5}$ (to
eliminate transients). The results for $L_{x,y,z}/L_0$
are plotted against $(\dot{\gamma}T_0)^{-1}$ in
\F\ref{fig:scalings}.
Linear least-squares fits to these data suggest scaling
exponents for $L_{x,y,z}$ of, respectively,
$-0.54\pm0.03$, $-0.65\pm0.03$, and $-0.60\pm0.04$
at the 95\% confidence level.
An alternative scaling, using the principal axes of the gradient
statistic \cite{WagnerYeomans99,stansell}, gives exponents for
$L_3, L_1, L_2$ of $-0.53\pm 0.04$, $-0.67\pm 0.03$, and $-0.64\pm 0.06$
(data not shown).
These results appear to rule out $L_y \sim \dot\gamma^{-3/4}$ which was
found in 2D \cite{stansell}. However, the range of Re accessible is
restricted to about 1 decade (260$\leq$ Re $\leq$ 2300); as in 2D, one
cannot rule out that these are effective exponents describing the
crossover region. These Re values are also high enough that a multiple
length scaling might be needed \cite{kendon2000}.

The quoted error margins do not, of course, allow for systematic error of which there are several sources (even discounting finite size effects), each at the likely level of several percent \cite{Kendon01,stansell}.
Accordingly these results do not rule out a common scaling of all three correlation lengths with a single exponent, $L_{x,y,z}/L_0 \sim (\dot{\gamma} T_0)^{-2/3}$, at least in the inertial limit of very large $(\dot\gamma T_0)^{-1}$ where the data hints that the three curves may saturate to fixed ratios. Conversely, the ever increasing difficulty to achieve NESS at small $(\dot\gamma T_0)^{-1}$ may point to a quite different behavior at low Reynolds
numbers. Suggestively, Fielding \cite{fielding} has recently performed 2D binary Stokes flow simulations finding no evidence of bulk NESS at Re $= 0$; this could
mean that inertia plays the role of a singular perturbation in this problem. Moreover, for a range of $\dot\gamma T_0$ around $10^{-3}$, NESS is easily achieved in 2D but not 3D: the ability to form connections in the vorticity direction might, at moderate and low Re, require formation of domains of extremely high aspect ratio before a NESS can be reached.
  
In conclusion, while open issues remain concerning the details of scaling and finite size behavior, our simulations present clear evidence for nonequilibrium steady states in 3D sheared binary fluids. The qualitative character of the NESS achieved in these simulations at high Re (low shear rate), which entails a balance between domain stretching under flow and coarsening driven by interfacial tension, strongly suggests that these results represent true bulk behavior. Since the effect of coarsening at fixed $\dot\gamma$ is to increase Re, indefinite coarsening \cite{Bray03} can seemingly be ruled out even at higher shear rates, although a different mechanism for achieving NESS may operate there.

\emph{Acknowledgements:} Work funded in part by EPSRC
EP/C536452 and GR/S10377. JCD acknowledges use of facilities  at the
Irish Centre for High-End Computing.
We thank S.~Fielding and R.~Adhikari for discussions.


\begin{thebibliography}{10}
\bibitem{SFM}M. E. Cates and M. R. Evans (Eds.), \emph{Soft and Fragile Matter, Nonequilibrium
Dynamics, Metastability and Flow}, IOP Publishing, Bristol 2000. 
\bibitem{Onuki02}A. Onuki, \emph{Phase Transition Dynamics}, Cambridge University Press, Cambridge
2002. 
\bibitem{Cates99}M.~E.~Cates, V.~M.~Kendon, P.~Bladon and J.-C.~Desplat, \emph{Faraday Disc.}
\textbf{112}, 1 (1999). 
\bibitem{WagnerYeomans99}A. J. Wagner and J.M.~Yeomans, \emph{Phys. Rev. E}
\textbf{59}, 4366 (1999); \emph{Phys. Rev. Lett.} \textbf{80}, 1429 (1998).  
\bibitem{Gonnella01}A. Lamura and G. Gonnella, \emph{Physica A} \textbf{294}, 295 (2001); F. Corberi, G. Gonnella and A. Lamura \emph{Phys. Rev. Lett.} \textbf{81}, 3852 (1998);
\emph{Phys. Rev. Lett.} \textbf{83}, 4057 (1999); \emph{Phys. Rev. E} \textbf{61},
6621 (2000); \emph{Phys. Rev. E} \textbf{62}, 8064 (2000). 
\bibitem{stansell}
P. Stansell, K. Stratford, J.-C. Desplat, R. Adhikari, and M. E. Cates
\textit{Phys. Rev. Lett.} \textbf{96} 085701 (2006).
\bibitem{catesmilner} M. E. Cates and S. T. Milner, \emph{Phys. Rev. Lett.} \textbf{62}, 1856 (1989).
\bibitem{Hashimoto888994}T. Hashimoto, T. Takebe and S. Suehiro, \emph{J. Chem. Phys.} \textbf{88}, 5875 (1988);
C. K. Chan, F. Perrot and D. Beysens, \emph{Phys. Rev. A} \textbf{43}, 1826 (1991);
A. H. Krall, J. V. Sengers and K. Hamano, \emph{Phys. Rev. Lett.} \textbf{69}, 1963
(1992); T. Hashimoto, K. Matsuzaka, E. Moses and A. Onuki, \emph{Phys. Rev. Lett.}
\textbf{74}, 126 (1995); Y. Takahashi, N. Kurashima, I. Noda and M. Doi, \emph{J. Rheol.} \textbf{38}, 699 (1994).
\bibitem{Bray03}
A.J. Bray, \emph{Phil. Trans. Roy. Soc. A} \textbf{361}, 781 (2003);
A. Cavagna, A.J. Bray,
and R.D.M. Travasso, \emph{Phys. Rev. E} \textbf{62}, 4702 (2000); A.
J. Bray and A. Cavagna, \emph{J. Phys. A} \textbf{33}, L305 (2000). 
\bibitem{Onuki97}A. Onuki, \emph{J. Phys. Cond. Mat.} \textbf{9}, 6119 (1997). 
\bibitem{DoiM91}M.~Doi and T.~ Ohta, \emph{J. Chem. Phys.} \textbf{95}, 1241 (1991). 
\bibitem{Kendon01}
V.M.~Kendon, et al.,
\emph{J. Fluid Mech.} \textbf{440}, 147 (2001);
V.M.~Kendon, J.-C.~Desplat, P.~Bladon and M.E.~Cates,
\emph{Phys. Rev. Lett.} \textbf{83}, 576 (1999);
I. Pagonabarraga, A.J.~Wagner, and M.E.~Cates,
\textit{J. Stat. Phys.} \textbf{107}, 39 (2002).
\bibitem{harting} J. Harting, J. Chin, M. Venturoli and P. V. Coveney, \emph{Phil. Trans. Roy. Soc. Lond.} \textbf{A 363}, 1895 (2005).
\bibitem{CatesPT}M. E. Cates, et al.,
\emph{Phil. Trans. Roy. Soc. A} \textbf{363}, 1917 (2005). 
\bibitem{Swift96}
M.~R.~Swift, E.~Orlandini, W.~R.~Osborn and J.~M.~Yeomans, \emph{Phys. Rev.
E} \textbf{54}, 5041 (1996). 
\bibitem{Adhikari05}
A. J. Wagner and I. Pagonabarraga, \emph{J. Stat. Phys.} \textbf{107}, 521 (2002). 
\bibitem{Bray94}A. J. Bray, \emph{Adv. Phys.} \textbf{43}, 357 (1994). 
\bibitem{d3q19} D. d'Humi\`eres, I. Ginzburg, M. Krafczyk, P. Lallemand and
L.-S. Luo,
\textit{Phil. Trans. R. Soc. A} \textbf{360}, 437 (2002).
\bibitem{jstatphys}
K. Stratford, R. Adhikari, I. Pagonabarraga, J.-C. Desplat,
\textit{J. Stat. Phys.}  \textbf{121}, 163 (2005).
\bibitem{supplementary} An animation of this run is available as
supplementary online material.
\bibitem{jury} S. I. Jury, P. Bladon, S. Krishna and M. E. Cates, \emph{Phys. Rev. E} \textbf{59}, R2535 (1999).

\bibitem{kendon2000}
V. M. Kendon,
\textit{Phys. Rev. E} \textbf{61} R6071 (2000).

\bibitem{fielding} S. M. Fielding, private communication.








\end{thebibliography}
\end{document}